# Analysis and Extraction of Tempo-Spatial Events in an Efficient Archival CDN with Emphasis on Telegram

*Melika Bahman-Abadi [a], M. B. Ghaznavi-Ghoushchi [b, 1]*

[a]Department of Computer Science, Shahed University, Tehran, Iran.
E-mail: melika.bahmanabadi@shahed.ac.ir

[b]Department of Electrical Engineering, Shahed University, Tehran, Iran.
E-mail: ghaznavi@shahed.ac.ir

**Abstract**

This paper presents an efficient archival framework for exploring and tracking cyberspace large-scale data called Tempo-Spatial Content Delivery Network (TS-CDN). Social media data streams are renewing in time and spatial dimensions. Various types of websites and social networks (i.e., channels, groups, pages, etc.) are considered spatial in cyberspace. Accurate analysis entails encompassing the bulk of data. In TS-CDN by applying the hash function on big data an efficient content delivery network is created. Using hash function rebuffs data redundancy and leads to conclude unique data archive in large-scale. This framework based on entered query allows for apparent monitoring and exploring data in tempo-spatial dimension based on TF-IDF score. Also by conformance from i18n standard, the Unicode problem has been dissolved. For evaluation of TS-CDN framework, a dataset from Telegram news channels from March 23, 2020 (1399-01-01), to September 21, 2020 (1399-06-31) on topics including *Coronavirus (COVID-19)*, *vaccine*, *school reopening*, *flood*, *earthquake*, *justice shares*, *petroleum*, and *quarantine* exploited. By applying hash on Telegram dataset in the mentioned time interval, a significant reduction in media files such as 39.8% for videos (from 79.5 GB to 47.8 GB), and 10% for images (from 4 GB to 3.6 GB) occurred. TS-CDN infrastructure in a web-based approach has been presented as a service-oriented system. Experiments conducted on enormous time series data, including different spatial dimensions (i.e., *Khabare Fouri*, *Khabarhaye Fouri*, *Akhbare Rouze Iran*, and *Akhbare Rasmi* Telegram news channels), demonstrate the efficiency and applicability of the implemented TS-CDN framework.

---

[1] Corresponding author:
M. B. Ghaznavi-Ghoushchi, Department of Electrical Engineering, Shahed University, Tehran, Iran.
E-mail address: ghaznavi@shahed.ac.ir
Tel.: +98-21-55212081.

## 1. Introduction

With the increasing scale of online news and social media, there is a growing interest in analyzing social issues and consumer trends. Researchers are actively working on detecting spatio-temporal event sentences in text data. However, not all events in a document are equally important for event analysis, and it is crucial to extract only the key events to improve the accuracy of event analysis(Kim, Yang, Park, & Jang, 2023). Commercial applications of sentiment analysis are being applied in marketing, politics, economics, health, and emergencies (Rodríguez-Ibánez, Casánez-Ventura, Castejón-Mateos, & Cuenca-Jiménez, 2023). Considering the semantic analysis of temporal data in tempo-spatial dimensions gives an obvious insight into events (Sachkov, Zhukov, Korablin, Raev, & Akimov, 2020). Behavioral analytics in cyberspace data can be used to explore and predict events. This is beneficial in deciding on critical issues and provides information that is not accessible and understandable without having a long-term data archive (Chen, Kong, & Mao, 2017; Huang & Wang, 2019).

Artificial intelligence and machine learning models can be utilized to analyze big data from social media platforms like Twitter and Facebook to identify and classify events as anomalous or not (Singh et al., 2022). Real-time responsible AI models for mental health analysis have the potential to improve mental health outcomes by enabling early detection, intervention, and support for individuals in need (Garg, 2023). In recent years, interest in instant messaging applications such as Telegram, WhatsApp, etc. has increased (Lareki, Altuna, Martínez de Morentin, & Amenabar, 2017). These messengers play a vital role in today's world in terms of the communication they provide (Asher, Caylor, & Neigel, 2018). The study of instant messaging networks is a new focus in research and has attracted widespread attention (Soares, Joia, Altieri, & Lander Regasso, 2021). The rise of social media platforms has revolutionized the way we consume news, but it has also led to the rapid spread of fake news (Kaliyar, Goswami, & Narang, 2021). The trends spread faster than the news (Khan, Nasir, Nasim, Shabbir, & Mahmood, 2021) so a neural network model of epidemics on social media has been created, given the epidemic of COVID-19 disease and its many economic and social impacts (Lymperopoulos, 2021). Massive online social networks contain hundreds of millions of nodes and often large amounts of information at the individual node level. This presents great opportunities for content recommendation and marketing (Nguyen, 2016). A deep learning model for stock forecasting that simulates buying and selling behaviors have been implemented due to (Ma, Han, & Wang, 2021) the users on social networks are usually affected by their first input data (Seo & Cho, 2021).

In a study, the probability of retweeting a message on Twitter was estimated using a classification method where, three groups of users were defined, including mass media sources, regular users, and opinion leaders (Camarero & San José, 2011). Tweets related to a specific event are also identified using the feel associated with each tweet. Semantic word analysis is a way to discover new events on Twitter (Sharma & Sharma, 2020). Events are identified according to their ontology and changing the feel associated with a tweet may make that event significant (Li, Nourbakhsh, Shah, & Liu, 2017). Besides, an incremental clustering approach has been designed to identify events in a Twitter data stream (Hasan, Orgun, & Schwitter, 2019) and a study has been conducted to track and distinguish events from heterogeneous news streams (Mele, Bahrainian, & Crestani, 2019). A semi-supervised system is also designed to help users automatically identify and visualize targeted events from Twitter (Hua, Chen, Zhao, Lu, & Ramakrishnan, 2013).

Tracking emotions towards different entities, recognizing different symbols of emotions, and extracting and ranking them in Twitter data are other interesting issues on how thinking and research have been formed (Giachanou, Mele, & Crestani, 2016). The challenges of verifying data on social networks include massive information, unverified information, rapid diffusion, altered posts, and so on. Due to the mentioned challenges a framework is proposed which incorporates post-wise features such as user-based, content-based, and lexical-based features, along with post sequences. The framework combines word embedding with bidirectional long short-term memory (BiLSTM) and utilizes a multilayer perceptron (MLP) to improve accuracy (Shelke & Attar, 2022).

Telegram as an instant messaging service, not only, has private messaging and encrypting messages, but also has strong and effective storage of data history and private cloud-based storage (Dargahi Nobari, Sarraf, Neshati, & Erfanian Daneshvar, 2021; Baumgartner, Zannettou, Squire, & Blackburn, 2020). The other affirmative reason to use Telegram is communication security, although another competitor WhatsApp dominates social media due to its simplicity (Sutikno, Handayani, Stiawan, Riyadi, & Much Ibnu Subroto, 2016) also message length characteristics and the number of @ per message are the most important features for detecting spam and non-advertising messages In Telegram. In particular, if the number of mentions in a message is more than 5, it can be a strong signal to detect spam messages (Dargahi Nobari, Reshadatmand, & Neshati, 2017).

## 2. Related works

Social Networks are the most extraordinary platform for acquiring data and detecting information. The COVID-19 global epidemic influences the population and mutates it. Usage of cyberspace for excavating data patterns leads to the survival of people from this phenomenon (Hou, Hou, & Cai, 2021). The problem of short text dispersion in social media makes semantic inference of temporal and spatial dimensions impractical. Nowadays tempo-spatial data modeling is significantly rising so an analysis of traffic data is done rely on event localization and geographical areas (Boghiu & Gîfu, 2020). A BiLSTM-based document classification model is proposed to identify important representative spatio-temporal event documents from a larger set of events. They create a training dataset of 10,000 gold-standard examples to train the model. Experimental results demonstrate that the proposed BiLSTM model outperforms the baseline CNN model, improving the F1 score by 2.6% and the accuracy by 4.5% (Kim et al., 2023).

Researchers have developed various methods to detect fake news, with sequential neural networks being commonly used to analyze the text sequence in a unidirectional manner (Kaliyar et al., 2021). A multi-modal framework called FR-Detect is implemented that combines user-related and content-related features to achieve high accuracy in fake news detection. The framework incorporates a sentence-level convolutional neural network to effectively combine publishers' features with latent textual content features (Jarrahi & Safari, 2023).

A semantic modeling approach invented by Kou et al. (2018), exploits the tempo-spatial specifications of the data stream in social networks. Monitoring geospatial topics in social media over time provide crucial information about natural disasters (Wu et al., 2020). In Japan, a harsh flood in 2019 led to the creating of flood detection software for crisis management level from the social media data stream (Shoyama, Cui, Hanashima, Sano, & Usuda, 2021) also the Kalman filter and related features extraction have been used to detect earthquake using a bulk of Twitter data (Sakaki, Okazaki, & Matsuo, 2010).

A framework was introduced to integrate big data semantically and detect time and space from various data resources (Santipantakis et al., 2020). Also, another framework is proposed to archive data from the Twitter data stream and then visualize the large-scale archive using Gephi (Groshek, de Mees, & Eschmann, 2020). The challenges in event detection from social media data include limited and noisy text, volume and velocity, data variety, data veracity, temporal information, event witness identification, multimodal event detection, event popularity prediction, rumor detection and event detection integration, and cross-platform and cross-language Detection. These challenges require advancements in natural language processing, text mining, information retrieval, social network analysis, and machine learning techniques to improve event detection from social media data (Li, Chao, Li, Lu, & Zhang, 2022).

With the mining of Instagram posts metadata, captions, and images, the user emotions are investigated and archived in a large-scale database for employment as a dictionary in sentiment analysis from the text in social media (Weismayer, Gunter, & Önder, 2021). Post-User Interaction Network (PSIN) framework models the interactions in a social context through a divide-and-conquer strategy. It decomposes the social context into three parts: post propagation tree, user social graph, and post-user interaction graph. Each part is processed individually, and then the information is integrated at the end (Min et al., 2022). A consumer review summarization model using natural language processing techniques and long short-term memory is developed. The model aims to provide businesses with summarized data to analyze consumer sentiment and make informed decisions. A hybrid feature extraction method, combining review-related and aspect-related features, is introduced to construct a distinctive feature vector for each review (Kaur & Sharma, 2023).

Evaluation of the above-mentioned researches unveils that the main target platforms are WhatsApp, Twitter, and Telegram. Among these three mentioned platforms, Telegram attracted fewer research interests. On the other hand, the conventional approaches suffer from the lack of combined tempo-spatial analysis. Therefore in this study, Telegram is selected as the target platform. Moreover, a tempo-spatial analysis approach is performed and a storage-efficient Content Delivery Network (CDN) is developed.

**3. Data modeling and analysis in tempo-spatial dimensions**

Cyberspace consists of the entities dissemination in tempo-spatial axes. A comprehensive look at the previous researches shows that today with the expansion of social networks in cyberspace, the evolution process of big data in the time axis is accruing as swiftly as possible. In this paper, since mining, exploring, and tracking multi-media events, a tempo-spatial model of the cyberspace data stream is investigated. Since the bulk of real-time data dissemination in social networks and websites, discovering and tracking data patterns is an interesting field, the more the data archive, the more exquisite prediction. The space dimension (i.e., cyberspace) refers to the diverse types of websites, social media, news channels, and newsgroups. The time dimension considered such as day, month, year, hour, minutes, and seconds.

A Content Delivery Network (CDN) is a group of geographically distributed data centers that are closer to the user and often are used to conservation of static content. This reduces the latency between fixed content and the user (Mokhtarian & Jacobsen, 2017). By utilizing CDN a copy of the content is stored in the content delivery network too (Ul Islam et al., 2019). Stored data on the CDN usually has an expiration date attribute called TTL (Time To Live). When TTL expires, then a specified service invokes to reloads data from the main data center to the local CDN (Scholl, Swanson, & Jausovec, 2019).

The collected data on tempo-spatial CDN (TS-CDN) gains to apply a diverse set of data modeling and indexing schemas. The cyber-temporal, cyberspace, and combined tempo-spatial data modeling, analysis, indexing, and data inquiry are now feasible utilizing TS-CDN. As indicated in Eq. (1), there are diverse versions of cyberspace $C$ at various time intervals. Each cyberspace version $c^{t_i}$ has a unique timestamp related to its occurrence time. Any modification in cyberspace content is denoted by $c^{t_k}$, $\{k = 1,2, ...\}$.

$$C = \langle c^{t_1}, c^{t_2}, ... \rangle \tag{1}$$

In Fig. 1, various types of files in each cyberspace are annotated. The target file types (i.e., images, documents, and videos) are represented in a colorful range of *images, documents, and videos*.

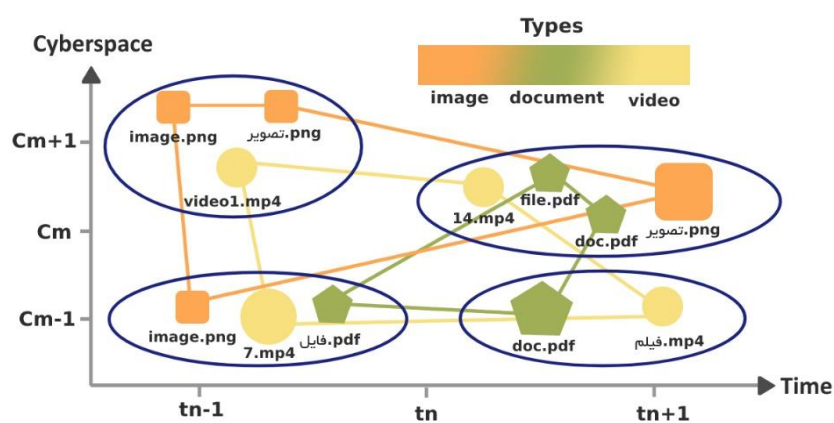

**Fig. 1. Data redundancy in cyberspaces.**

Cyberspace $C_{m-1} = [\{image.png, 7.mp4, فایل.pdf\}, \{doc.pdf, فیلم.mp4\}]$ in variant time intervals, cyberspace $C_m = \{14.mp4, file.pdf, doc.pdf, تصویر.png\}$ and cyberspace $C_{m+1} = \{imgae.png, video1.mp4, تصویر.png\}$ are depicted in Fig. 1. There are duplicate files with dissimilar names (i.e., non-English like Persian تصویر) in cyberspaces. For instance in Fig. 1, *video1.mp4*, in cyberspace $C_{m+1}$ exists in cyberspace $C_m$ and $C_{m-1}$ as *14.mp4* and فیلم*.mp4*, respectively.

In this paper to avoiding redundancy in archiving data, file content digesting is utilized. In section 4 the procedure of creating an efficient tempo-spatial CDN in detail is explained. In section 5 the achieved results unveil the efficiency of TS-CDN.

**4. Creating an efficient tempo-spatial content delivery network**

On Twitter, the users can send a message from one account to other accounts. They retweet messages unlimitedly (Rao, Vemprala, Akello, & Valecha, 2020). In instant messengers (i.e., Telegram) sharing information is done through forwarding messages from one channel to other groups or contacts list (Dargahi Nobari, Sarraf, Neshati, & Erfanian Daneshvar, 2021). There are duplicate contents in instant messenger data. For avoiding redundancy in archiving of data stream especially huge size file, utilization an approach that stores unique data is essential. Besides, archived files have lots of problems such as non-standard and lengthy naming, incompatibility with charsets, conflict with the character size limitation of guest or host operating system, multi-lingual naming, and max-length restriction of string (text field) in the databases. In this paper to produce an efficient archive of the Telegram data stream and to deal with these challenges, Tempo-Spatial Content Delivery Network (TS-CDN) is implemented. As illustrated in Fig. 2, the process of creating an effective CDN of the multimedia data stream has seven stages.

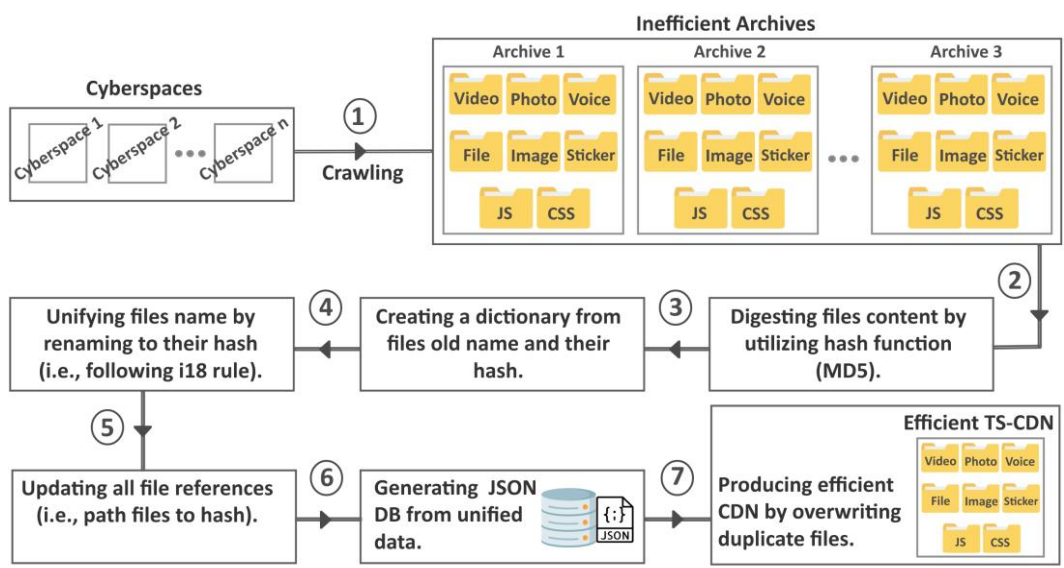

**Fig. 2. The process of creating an effective CDN from the Telegram.**

The first fold is crawling data streams from cyberspace. The second fold is using content digest techniques for labeling files name. Between SHA256 and MD5 options, MD5 is preferred due to less complexity and faster processing rate. Unifying files name by files content hash, resolved the issue of files with the same content but different file names. Therefore, duplicate files are resolved in TS-CDN. The third fold is producing a dictionary from the old path address of files and their new name (i.e., hash). This is required for updating references in the fifth fold. The fourth fold is renaming all kinds of files to their content hash. TS-CDN follows the i18 internationalization rule and supports Unicode. It is implemented as a cross-platform, without OS dependency. In the proposed system, if a file with a Persian name is uploaded or download; all are seamlessly converted to 32 hashed characters. The fifth fold is updating all file references. Altering the path address of files is essential due to changing their original name to the hash. The sixth fold is creating a JSON database of text messages, messages date, images, and videos for analyzing in tempo-spatial dimensions. The last fold is producing an efficient archival CDN from hashed data by overwriting duplicate files and updating the master index.

Using CDN not only reduce the bulk of the archived data significantly but also in turn is gainful for an integrated unique long-term archive from cyberspace for temporal analytics and estimation. Creating an effective resumable archive in TS-CDN means that distributed data are integrated with unique names related to original contents. So choosing desired cyberspace with any time interval is possible for creating an efficient archive. The process of creating an effective CDN from Telegram is summarized in Algorithm 1.

```
Algorithm 1 Creating CDN from Telegram Archive
1: input = html file
2: while read a html H
3:     while read a hyperLink L in H
4:         hash = HashContent(L)
5:         Create Dictionary(fileName,hash)
6:         Internationalization by fileName ← hash
7:         Create JSONDB(message,media,date)
8: Merge mainCDN ← allCDN
```

## 5. Investigation of content distribution in TS-CDN

In this section, the exact investigation of archived resources such as volume, duration, and number by considering their extensions are performed. The number of news posts, images, images after the hash, videos, and videos after the hash, files with the English names, and files with Persian names in the archive are indicated in Fig. 3 by applying the i18n standard.

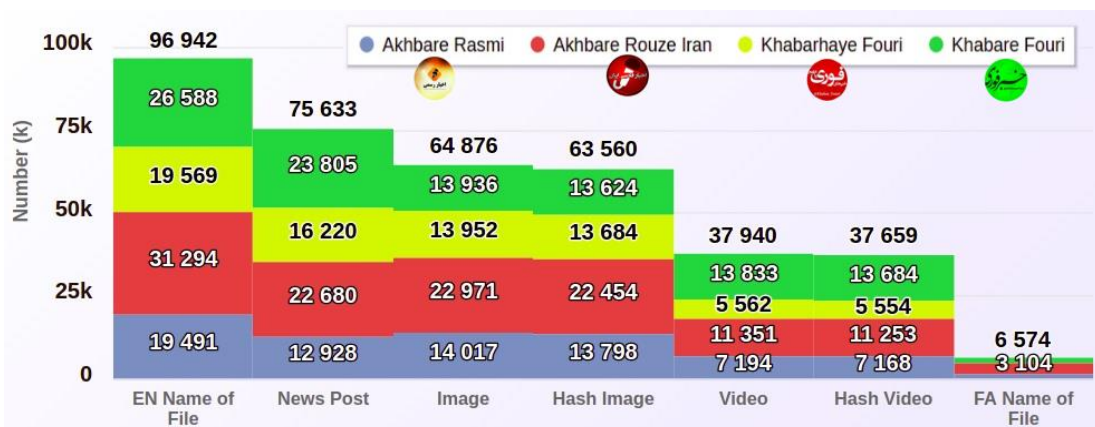

**Fig. 3. The number of news posts, images, images after the hash, videos, videos after the hash, files with English names, and files with Persian names in the archive.**

In Fig.3 the largest number of photos is related to the *Akhbare Rouze Iran* archive, and the total number of images and videos in the *Akhbare Rouze Iran* channel archive is more than other archives and is equal to 34,332. Fig. 4 displays the number of files, default icons, voice messages, CSS, JS, and stickers for each channel. According to Fig. 4, the *Khabare Fouri* channel with 414 files has the biggest number of files than other channels. Also, the only channel with stickers is the *Akhbare Rasmi* channel with 16 stickers. The largest volumes of archives belong to *Khabare Fouri* (30.4 GB), *Akhbare Rouze Iran* (28.p GB), *Akhbare Rasmi* (15.7 GB), and *Khabarhaye Fouri* (11.2 GB), respectively.

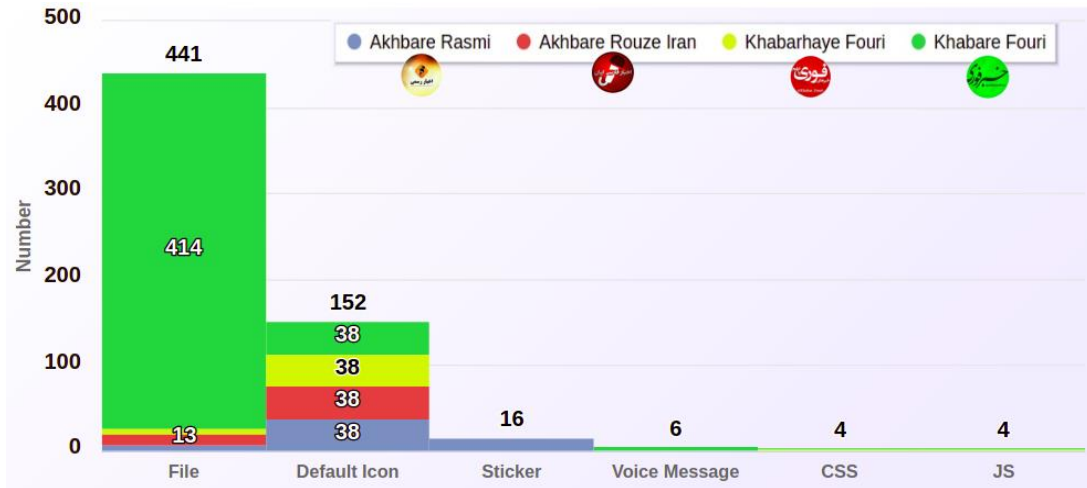

**Fig. 4. The number of other channel archive resources.**

Using hash algorithms is leads to reducing the archive volume. After renaming all the files in the channel archive to their hash content, the duplicate files will replace, and the size of the media files will significantly reduce. This conclusion is summarized in Table 1.

**Table 1**
**The number of items before merged into TS-CDN and after merged into TS-CDN.**

| News Channel | Number of items before merged into TS-CDN | | Number of items after merged into TS-CDN | |
|---|---|---|---|---|
| | Video | Image | Video | Image |
| Khabare Fouri | 13833 | 13936 | 13684 | 13624 |
| Khabarhaye Fouri | 5562 | 13952 | 5554 | 13684 |
| Akhbare Rouze Iran | 11351 | 22971 | 11253 | 22454 |
| Akhbare Rasmi | 7194 | 14017 | 7168 | 13798 |

Fig. 5 represents the number of different files when locating in TS-CDN. Table 2 unveils the efficiency of the TS-CDN storage mechanism regarding archive files and

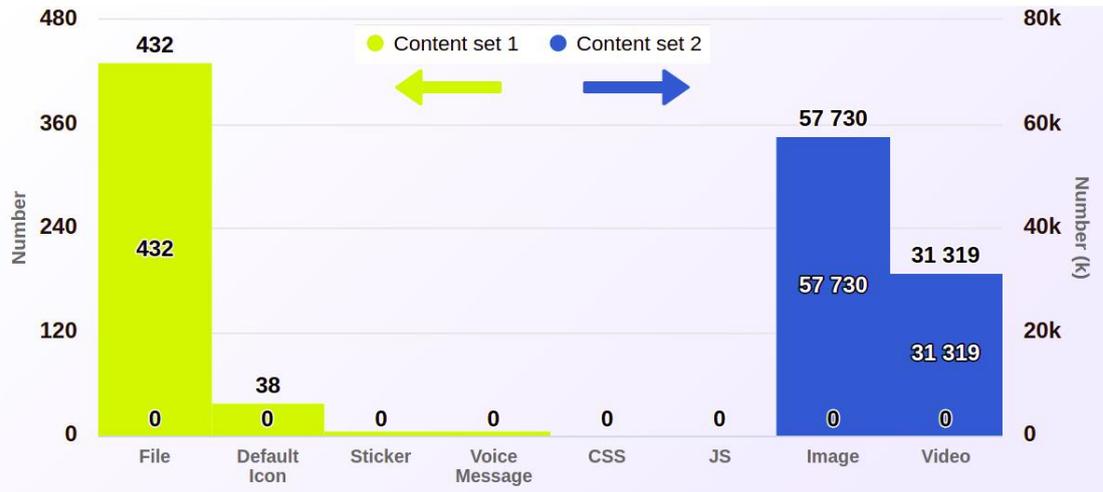
Fig. 5. The number of different files in TS-CDN.

media such as a 38.9% decrease for videos (from 79.5 GB to 47.8 GB) and 10% for images (from 4 GB to 3.6 GB).

Table 2
The comparison of file volumes before merged into TS-CDN and after merged into TS-CDN.

| Files | Volume before merged into TS-CDN (GB) | Volume after merged into TS-CDN (GB) | Decrease percentage |
|---|---|---|---|
| **Videos** | 79.5 | 47.8 | 39.8 |
| **Images** | 4 | 3.6 | 10 |
| **CSS & JS files** | $6*10^{-5}$ | $3*10^{-5}$ | 50 |
| **Miscellaneous files** | 2.6 | 2.5 | 3.8 |

So far, the archive files, the archive volume of each news source, CDN concepts, and the necessity to use TS-CDN are considered. In sections 5.1, 5.2, and 5.3 the archival sources, volume, duration of the media files, and cyberspace ranking in detail, are discussed and they are analyzed from a statistical point of view.

**5.1 Investigation of media files by format types**

In this stage, the analysis of files based on format types is discussed. As represented in Fig. 6, the duration of media files by file format and newsgroups is visualized. The .mp4 video format is the dominant multimedia format.

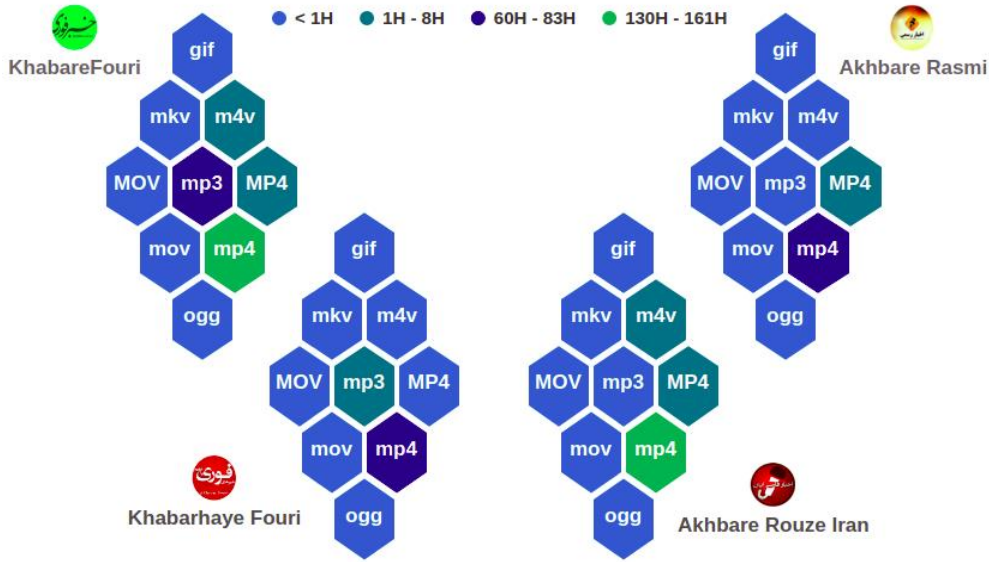

Fig. 6. Duration of media files in TS-CDN archive.

As demonstrated in Fig. 7, the size of video files, audio files, and images in multiple formats are illustrated. In this chart, the accumulated file size of the MP4 format is dominant with 81.61 GB which belongs to the *Khabare Fouri* channel.

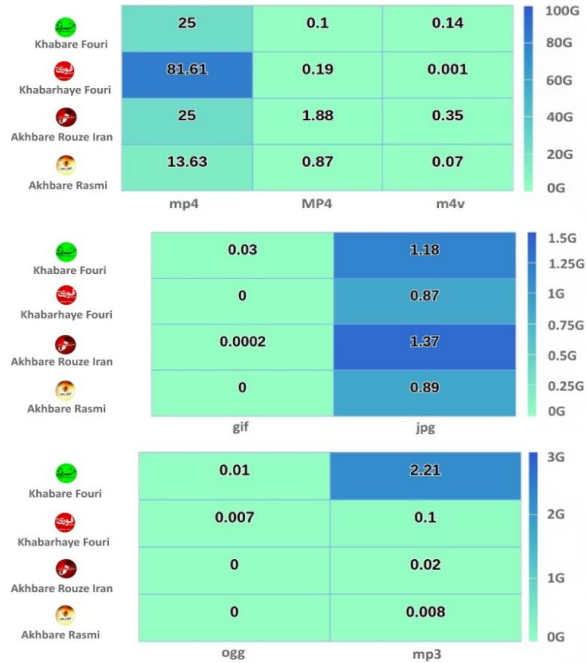

Fig. 7. Media volumes based on format types.

**5.2 Number of media files by size**

    In this section, a general investigation about the number of files in terms of different sizes is performed. Fig. 8 reveals the number of files based on variant sizes. Most files are less than 10k in size and belong to the *Akhbare Rasmi* channel. The highest number is related to video files with 79 MB volumes. The lowest number is related to photo files with 4 KB volume.

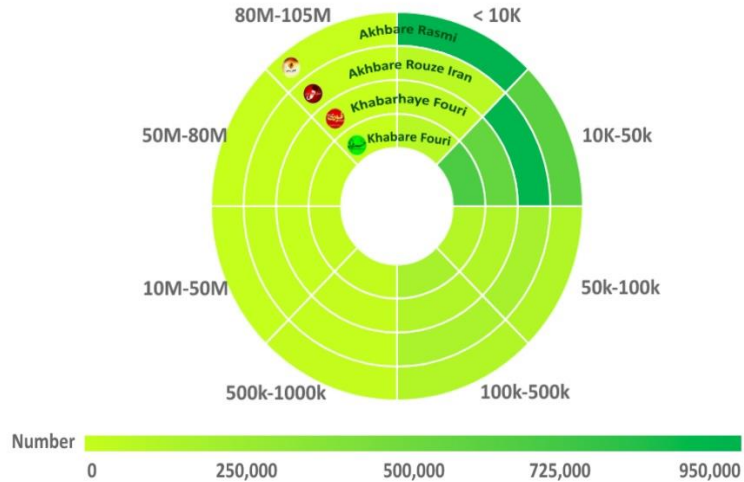

**Fig. 8. The frequency of files based on distinct volumes.**

**5.3 News channel rankings comparison**

    In the last part, the ranking of news channels based on the number of news and the total number of media files are presented. Towards this objective, the total number of news posts by news channels is extracted. According to Fig. 9, the maximum number of posts is related to *Khabarhaye Fouri*. The number of file types is also specified via their format in vertical axes.

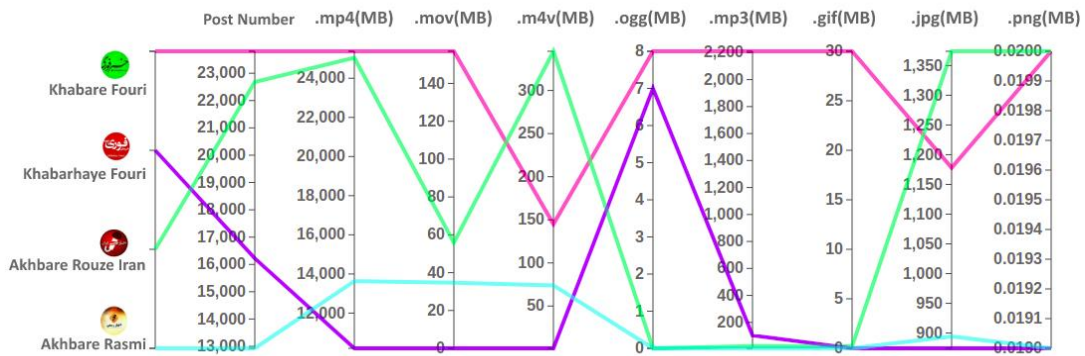

**Fig. 9. News channel rankings comparison.**

## 6. Text processing in tempo-spatial dimensions

Concerning the definition given in section 3 about cyberspace and cyber-temporal concepts, in this section, we aim to describe the time-travel text search, Term Frequency-Inverse Document Frequency (TF-IDF), time-travel inverted index, and temporal coalescing. To this end, first, the time-travel text search and then the term-based and time-based text processing is described.

### 6.1. Time-travel text search

The temporal update is vital in the evaluation and investigation of time series and real-time data streams. This update is a key issue in regularly update streams including news, microblogs, twits, and posts (Gao, Wang, Li, Shao, & Song, 2017). A general schema on modeling temporally annotated events is depicted in Fig. 10. As illustrated in Fig. 10, each classification includes a set of occurrences called cyberspace ensembles. Each cyberspace ensemble contains different kinds of events and sub-events. The events and sub-events are displayed based on time and Term Frequency-Inverse Document Frequency (TF-IDF) score in the x and y axes, respectively. The consequence of time points $t_1, t_2, t_3, \ldots, t_{n+1}$ is indicating the temporal trigger points.

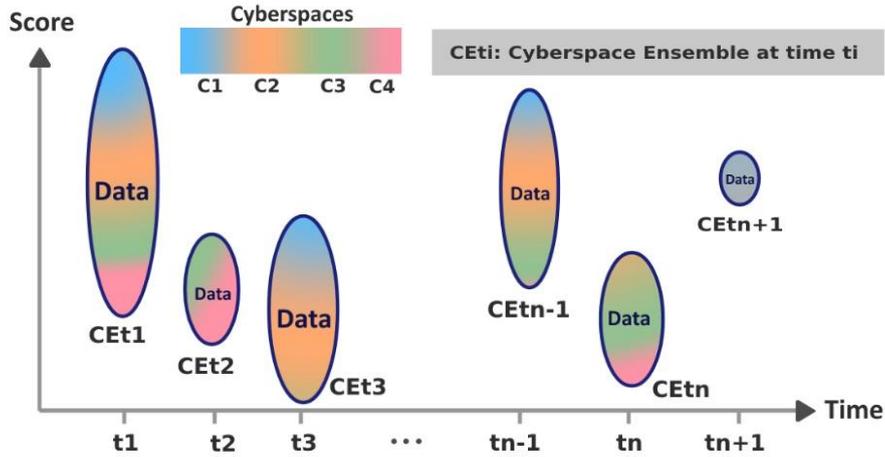

Fig. 10. Cyberspace ensembles in social networks.

In Fig. 11, events and sub-events in each cyberspace ensemble are annotated. The target cyberspaces (i.e., channels, groups, social networks, and web pages) are represented in a colorful range of $C_1, C_2, C_3,$ and $C_4$. Different types of events and sub-events are remarked in circles, triangles, and squares as depicted in Fig. 11.

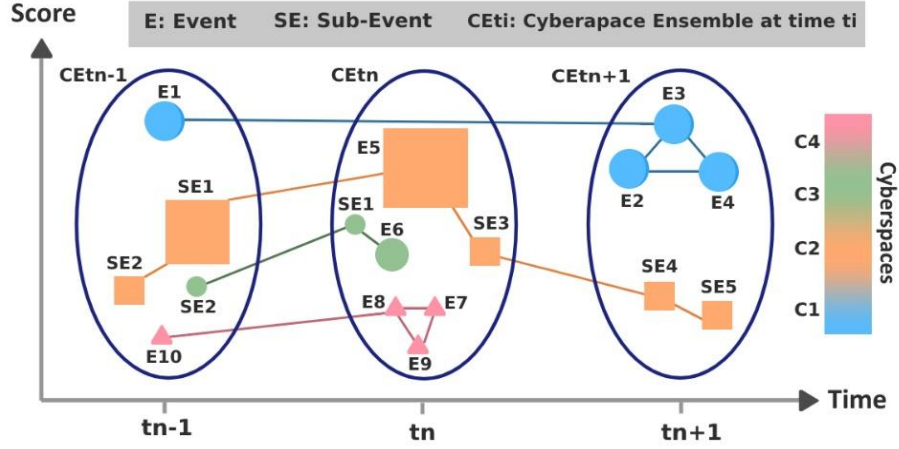

**Fig. 11. Cyberspace ensembles details.**

Each event may have multi sub-events and occur in some cyberspace ensembles. For instance, event $E_{blue} = \{E_1, E_2, E_3, E_4\}$ is originated from one cyberspace and the same context. This event has occurred at time sequences of $t_{n-1}$ and $t_{n+1}$, but $t_n$. The pink triangles represent another event $E_{pink} = \{E_7, E_8, E_9, E_{10}\}$ has identical incidents at $t_{n-1}$ and $t_n$ with the most occurrence at $t_n$. The event and sub-events of event $E_{orange} = \{E_5, SE_1, SE_2, SE_3, SE_4, SE_5\}$ are distributed at $t_{n-1}$, $t_n$, and $t_{n+1}$ with most at $t_n$. The event and sub-events of event $E_{green} = \{E_6, SE_1, SE_2\}$ are occurred at $t_{n-1}$, and $t_n$ with most at $t_n$. Each one of the main ovals $O_{white} = \{CEt_{n-1}, CEt_n, CEt_{n+1}\}$ is a set of classified data through applying the utilized algorithms in TS-CDN. These algorithms make the events explicit in tempo-spatial dimensions and make the favorite news easy to follow.

As indicated in Eq. (2), there are diverse versions of event $E$ at distinct time intervals. Each event version $e^{t_i}$ has a unique timestamp related to its occurrence time. Any modification in event is denoted by $e^{t_j}$, $\{j = 1, 2, ...\}$.

$$E = \langle e^{t_1}, e^{t_2}, ... \rangle \tag{2}$$

Eq. (3) demonstrates that valid time interval $val(e^{t_i})$ of version $e^{t_i}$ is $[t_i, t_{i+1})$ if a newer version with associated timestamp $t_{i+1}$ exists. If there is not any newer version, the valid time interval equals to $[t_i, now)$ where "$now$" is supposed to be the current time and there is no timestamp greater than "$now$".

$$val(e^{t_i}) = \begin{cases} [t_i, t_{i+1}) : & e^{i+1} \in e \\ [t_i, now) : & otherwise \end{cases} \tag{3}$$

Also with remarking $t_b$ and $t_e$ as the beginning and end times, respectively the time-based query $q^{[t_b, t_e]}$ is consists of two folds: One fold is the keyword-based query, and the other fold is either time interval $[t_b, t_e]$ or time point queries.

As discussed in this section, tracking events renewal in the time axis is useful for extracting temporal changes of data. Extending the data exploration to diverse kinds of cyberspaces ensures to have designated events and sub-events clarification. All events and sub-events are produced in cyberspace and cyber-time axes and are categorized based on semantic score. This score is calculated via tempo-spatial TF-IDF. In the next section, the tempo-spatial TF-IDF procedure will be described.

**6.2 Utilizing term frequency-inverse event frequency in Tempo-Spatial CDN**

In data retrieval, Term Frequency-Inverse Document Frequency (TF-IDF) is a weighting factor that is used to explore the importance of a term in context. In Tempo-Spatial CDN, documents are replaced and modeled as events which are called here as Inverse Event Frequency (IEF). So, IEF is the map of IDF utilizing tempo-spatial events. IEF reduces the weight of the two phrases groups. First, terms that often occur in the event. Second, terms that their absence from the event does not affect information retrieval. IEF increases the weight of meaningful terms that occur less frequently. IEF is according to Eq. (4) where, $|E|$ is the number of events and $ef_v$ is the number of events that have the word frequency $v$.

$$IEF_v = \log \frac{|E|}{ef_v} \tag{4}$$

Thus TF-IEF consists of the product of TF and IEF according to Eq. (5) where $TF_{v,e}$ is the number of term $v$ duplication in event e.

$$TF\text{-}IEF = TF_{v,e} * IEF_v \tag{5}$$

The vector space model in data retrieval consists of textual and query expressions weighted vectors. These weights have initialized for the number of expressions in one or more textual document sets. In Eq. (6) for each event and query, the angle cosine between the two vectors is given. The larger the cosine value, the greater the correlation between the query and event.

$$\cos(\vec{q}, \vec{e}) = \frac{\vec{q} \cdot \vec{e}}{|\vec{q}| \cdot |\vec{e}|} \tag{6}$$

In the TS-CDN system, the related events are returned based on the user query. The query phrase includes two sections. One is, keywords and the other is time (i.e., date and 24 hours a day format). TS-CDN can extract news trends at any time interval and display them to the user in an interactive manner. This paper outlines temporal coalescing and semantically correlation of collected items by time-travel inquiry. The process of calculating TF-IEF and adapting news events with the query is presented in Algorithms 2 and 3, respectively.

**Algorithm 2** TF-IEF
1: score for term t in event $e = TF(t,e) *$ IEF $(t)$
2: where
3: $IEF =$ Inverse Event Frequency
4: $TF =$ Term Frequency
5: $TF(t,e) = \frac{Term\ t\ frequency\ in\ event\ e}{Total\ words\ in\ event\ e}$
6: $IEF(t) = \log_2 \left( \frac{Total\ events}{events\ with\ term\ t} \right)$
7: and
8: $t =$ Term
9: $e =$ Event

**Algorithm 3** Adaptation between query and events
1: procedure Adaptation($event$)
2:     tokens = tokenize($eventText$)
3:     **for** all $tk \in$ tokens **do**
4:         tk = removeStopWord($tk$)
5:         tk = stemming($tk$)
6:         score = TF-IEF($tk$)
7:         wordVector.put($tk, score$)
8:     categoriesVecotrs = loadCategoryVectors()
9:     **for** all ca in categoriesVecotrs **do**
10:        similarity = cosineSimilarity ($ca, wordVector$)
11:        adaptation.put($ca.name, similarity$)
12:     save adaptation

As stated in this section, TS-CDN uses TF-IEF semantic algorithm that calculates the term duplications in events. This term is the phrase query that the user entered into the TS-CDN system for tracking news trends. In this procedure, the similarity between query and events is computed. In the next section, the inverted index data structure will be expressed.

### 6.3 Time-traveled inverted index

The inverted index is one of the standard methods of text indexing. It is a type of data structure that stores the mapping of content (i.e., words) to the occurred place of the event text. This data type enables TS-CDN to pinpoint the target tempo-spatial points. It is used to accelerate the search performance and consists of two main components: words and words incident. For each word in events; the event's name that includes word is stored in this data structure. The process of creating an inverted index is presented in Algorithm 4. The posting list has a data structure like a typical list which has three components: First, is the name of the event. The second is the number of word duplications in that event. The third is the number of indexes in the event text in which the word appears.

---
**Algorithm 4** Tempo-Spatial Inverted Index
1: input = event
2: output = res (search results)
3: **while** *read an event E*
4:    **while** *read a term T in E*
5:       **if** $find(dictionary, T) == false$ **then**
6:          Insert(dictionary,T)
7:       Get T's posting list.
8:       insert a node to T's posting list.
9: write the inverted index to disk.

---

### 6.4 Temporal coalescing of related events in TS-CDN

As explained in section 6.3 for each event version and its supporting expressions in an event context, a posting list is created. Duplicate expressions and regular updating of events lead to a huge number of lists. Often alternations to the event text may be minor (i.e., spelling changes) therefore a lot of redundancy between sequential versions of an event is possible. In this section, the technique that eliminates redundancy and is effective in reducing the number of posting lists is described. Towards this objective, the posting lists with time adjacency and including different versions of an event (i.e., identical TF-IEF score) are merged and considered as one event. Eq. (7) indicates that cyberspace ensembles consist of $n$ sequential time-correlated posting lists. Eq. (7) determines the existence of a particular term in event $e$ with term repetition $r$ at different time intervals.

$$M = <(e,[t_1,t_2),r_1)\ldots(e,[t_i,t_{i+1}),r_i)> \qquad (7)$$

After merging the posting list the consequence in Eq. (8) is created. It consists of $|C| = j \leq i$ coalesced posting and includes the same time interval as the main posting $M$ so that $t_1 = t_1'$ and $t_{i+1} = t_{j+1}'$.

$$C = < (e, [t_1', t_2'], r_1') \ldots (e, [t_j', t_{j+1}'], r_j') > \tag{8}$$

Fig. 12 demonstrates a sequence of non-integrated events in the tempo-spatial dimensions with a span of different scores as mentioned in Eq. (7). In Fig. 13, the events are put together which have time adjacency and almost identical scores as each other. Finally, the output of integrated events in the tempo-spatial dimensions is represented

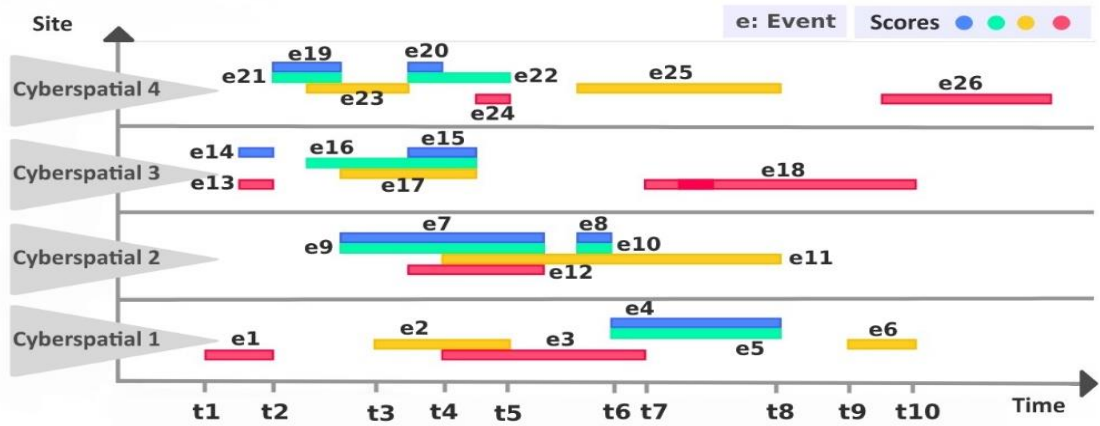

**Fig. 12. Unintegrated events in tempo-spatial dimensions.**

in Fig. 13. It depicts the stated temporal coalescing in Eq. (8).

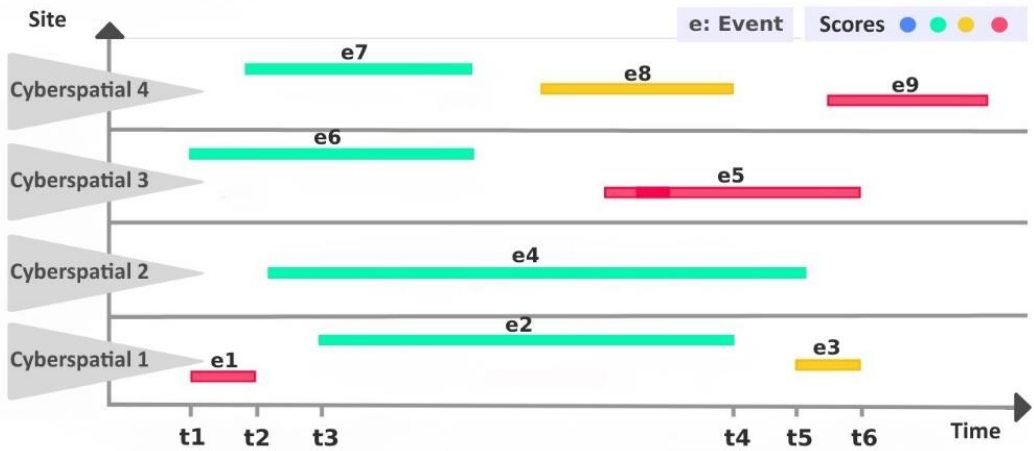

**Fig. 13. Integrated or coalesced events in tempo-spatial dimensions.**

As pointed out so far, TF-IEF is a measure for detecting frequency and emphasis of terms in events. Merging events that have time proximity and are in the same range of TF-IEF score decreases redundancy of diverse versions of an event. Using tempo-spatial inverted index accelerates detection of events and sub-events related to an inquiry. By segmenting each block of data via semantic syntax and natural language processing, meaningful events that are separable will extract. So a model for detecting trends that including sematic and time coalescing is created. This model supports time-travel and term-based queries. In the following TS-CDN functionality in deep will be represented.

**7. Tempo-spatial analysis and exploration in the news archive**

In this section, analyzing the archive of news in cyberspace is done. For this research, data is provided by crawling and archiving Telegram. In this paper a crawled dataset of Telegram news channels such as *Khabare Fouri*, *Khabarhaye Fouri*, *Akhbare Rouze Iran*, and *Akhbare Rasmi* from March 23, 2020 (1399-01-01), to September 21, 2020 (1399-06-31) is provided. In the TS-CDN system by searching a phrasal inquiry, related news is returned. Then channel activity is monitored and tracked in that specific domain. In Fig. 14, the continuity of events in all 4 channels regarding *Coronavirus (COVID-19)* outbreak news during the specified period is illustrated. The continuity between events in every four channels indicates the popularity of *Coronavirus (COVID-19)* news.

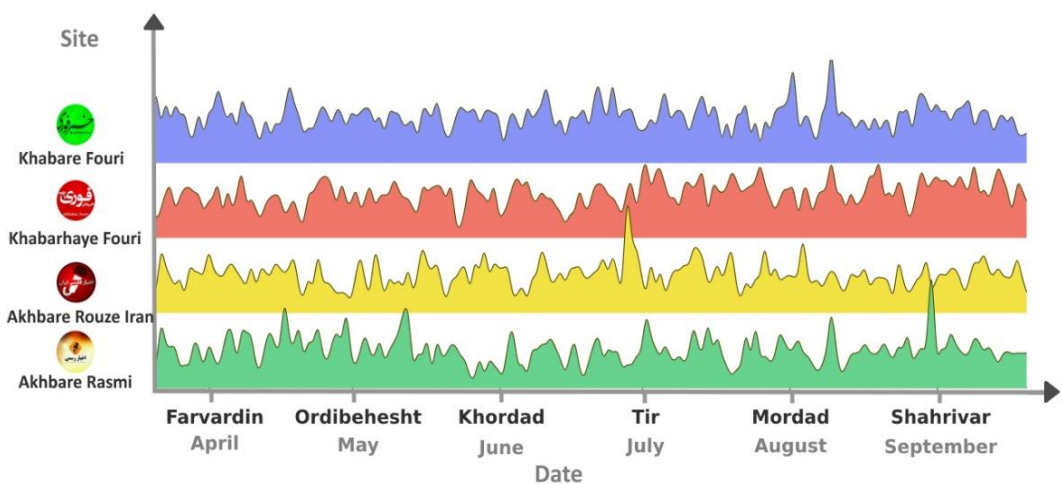

**Fig. 14.** Coronavirus news events in tempo-spatial dimensions in 2020 (1399).

In Fig. 14, if each month is separated by a hypothetical line, then almost four peaks for each channel are visible. In Table 3, the number of *Coronavirus (COVID-19)* news on weekends is summarized. According to Table 3, the number of *Coronavirus (COVID-19)* events on Iran weekends (i.e., Thursday and Friday) is more rather than the previous day (i.e., Wednesday) and the next day (i.e., Saturday). So each peak in Fig 14, can be considered as the weekend.

**Table 3**
The number of Coronavirus (COVID-19) news events in a weekend window during the period from March 23, 2020 (1399-01-01), to September 21, 2020 (1399-06-31).

| Months | Khabare Fouri | | | Khabarhaye Fouri | | | Akhbare Rouze Iran | | | Akhbare Rasmi | | |
|---|---|---|---|---|---|---|---|---|---|---|---|---|
| | Wed | Thu-Fri | Sat | Wed | Thu-Fri | Sat | Wed | Thu-Fri | Sat | Wed | Thu-Fri | Sat |
| April (Farvardin) | 245 | 552 | 384 | 269 | 546 | 320 | 194 | 459 | 281 | 98 | 269 | 145 |
| May (Ordibehesht) | 153 | 285 | 168 | 146 | 293 | 185 | 150 | 238 | 174 | 85 | 165 | 99 |
| June (Khordad) | 131 | 239 | 153 | 136 | 239 | 141 | 91 | 158 | 102 | 80 | 151 | 96 |
| July (Tir) | 177 | 283 | 217 | 109 | 170 | 140 | 113 | 132 | 144 | 98 | 134 | 137 |
| August (Mordad) | 151 | 222 | 158 | 120 | 164 | 112 | 118 | 189 | 140 | 99 | 144 | 94 |
| September (Shahrivar) | 80 | 164 | 102 | 71 | 146 | 86 | 82 | 111 | 74 | 59 | 95 | 63 |
| Total | 937 | 1745 | 1182 | 851 | 1558 | 984 | 748 | 1287 | 915 | 519 | 958 | 634 |

According to Table 1, the weekend activities unveils that *Khabare Fouri* comes with 82%, 35%, 12% more than *Akhbare Rasmi*, *Akhbare Rouze Iran*, and *Khabarhaye Fouri*, respectively. Fig. 15 indicates the *vaccine* news events over March 23, 2020 (1399-01-01), to September 21, 2020 (1399-06-31). The density of events in September (Shahrivar) reveals that the co-occurrence of events about the *vaccine* in Iran.

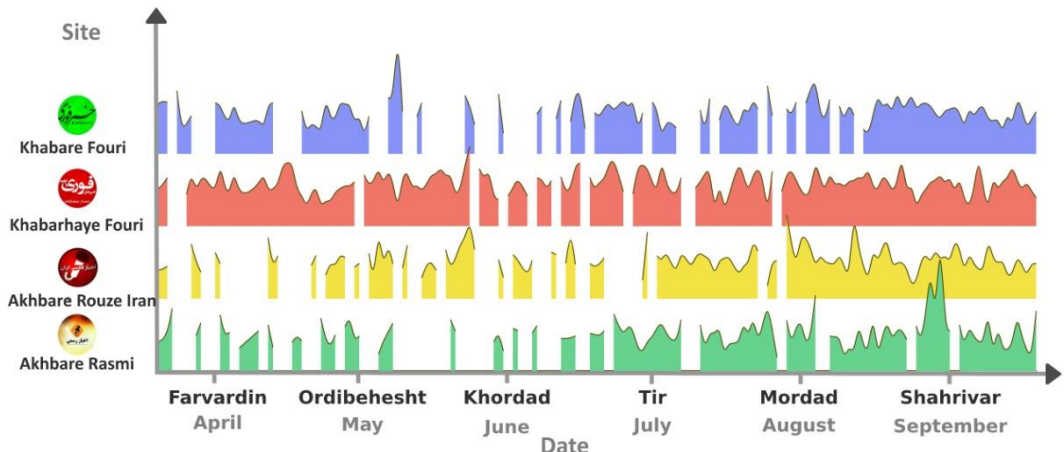

**Fig. 15. Vaccine news events in tempo-spatial dimensions in 2020 (1399).**

As depicted in Fig. 16 *schools reopening* in April (Farvardin) and September (Shahrivar) are more noteworthy than other months. After the Nowruz holiday in April (Farvardin) and ending summer in September (Shahrivar) *schools reopening* is a viral topic in Iran. So the data stream in these two months is not ignorable. Fig. 17 demonstrates that *Khabare Fouri* has more activity in *earthquake* disasters.

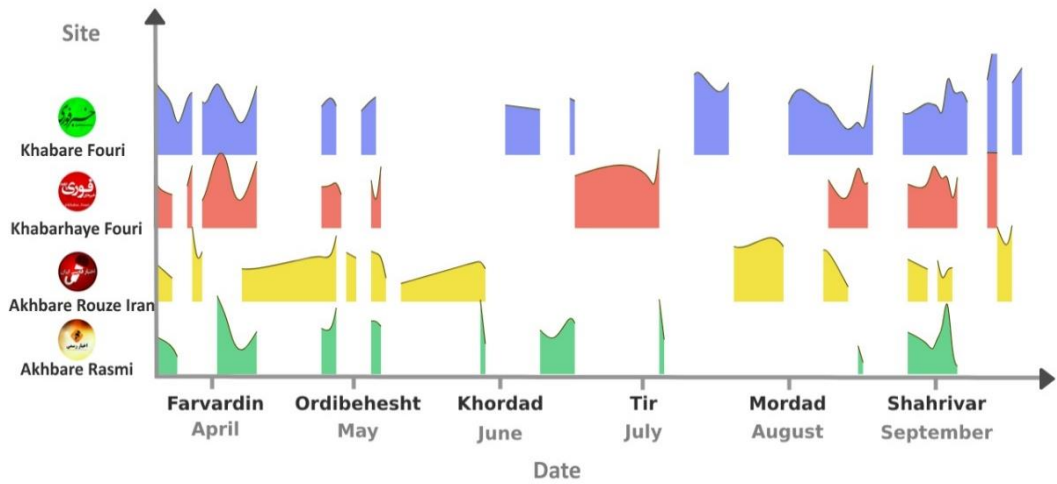

**Fig. 16.** Reopening schools news events in tempo-spatial dimensions in 2020 (1399).

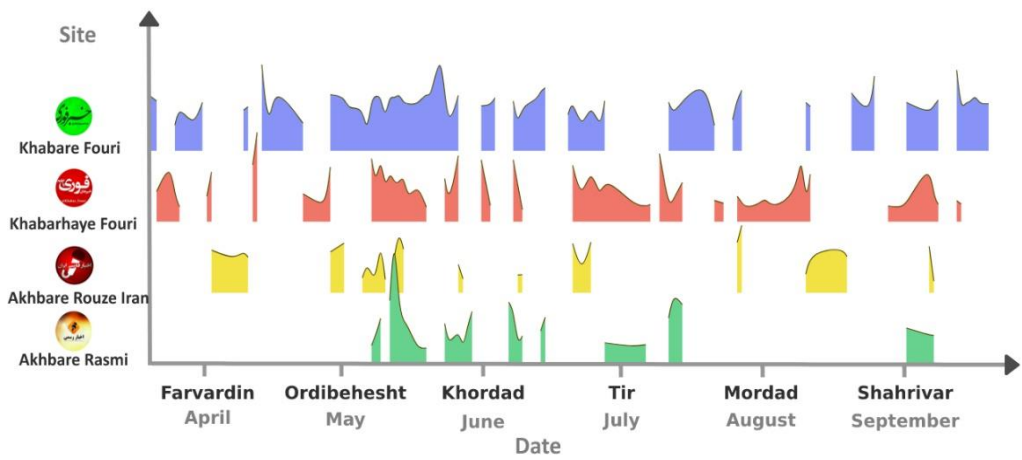

**Fig. 17.** Earthquake news events in tempo-spatial dimensions in 2020 (1399).

As shown in Fig. 18 *fire* events in June (Khordad) and July (Tir) have more occurrences. Therefore, it can be concluded that due to hot weather and drying of meadows, forest, and rangelands fire may occur more often.

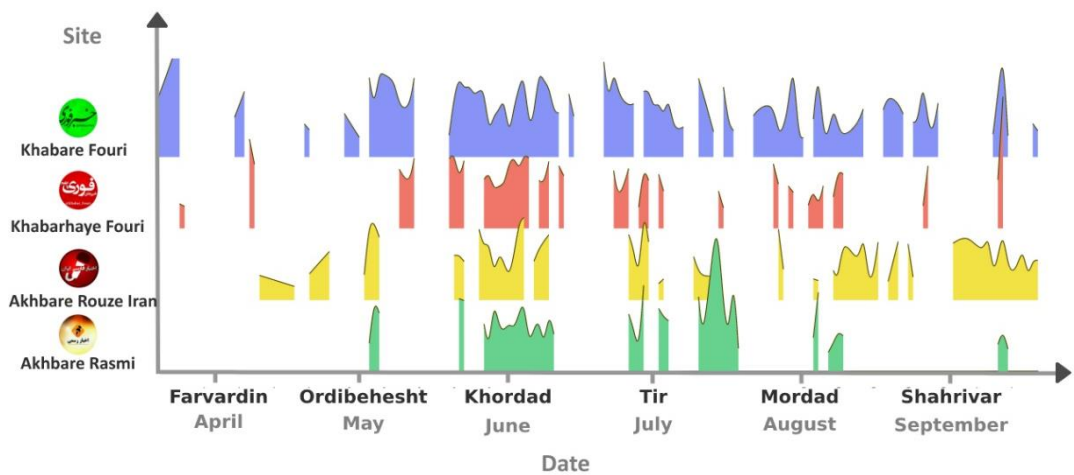

**Fig. 18.** Fire news events in tempo-spatial dimensions in 2020 (1399).

In these graphs, the TF-IEF score of the news is evaluated. The association between time-based slots, explains the conclusion of this study. For instance, the flood in some seasons such as spring and fall is more frequent. Fig. 19 indicates the integrity between *flood* news is more sensible in April (Farvardin), May (Ordibehesht), and June (Khordad) in the *Khabare Fouri* channel. Also the coalescing of events in this time interval is more observable.

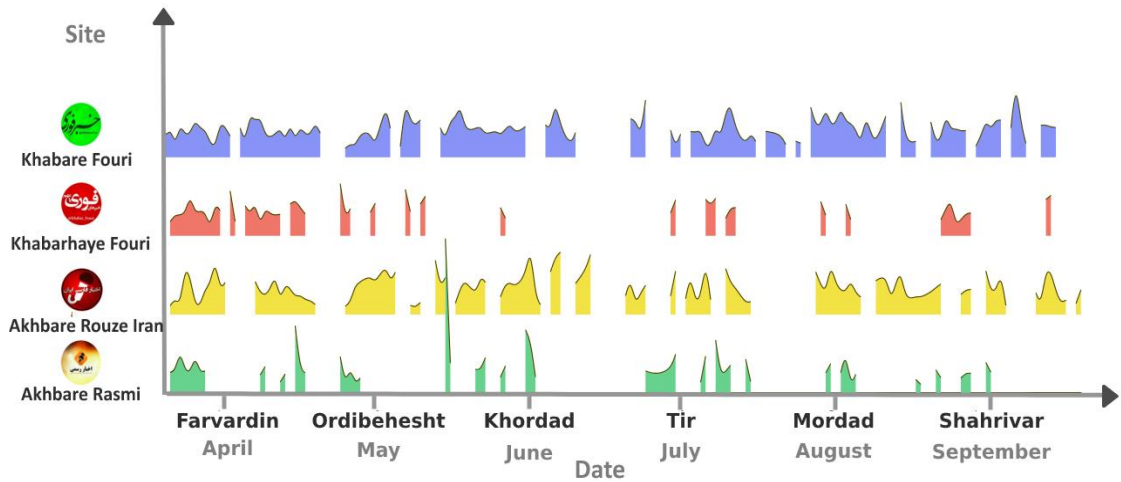

**Fig. 19. Flood news events in tempo-spatial dimensions in 2020 (1399).**

Fig. 20 indicates that *justice shares* (Sahame Adalat) events distribution from March 23, 2020 (1399-01-01), to September 21, 2020 (1399-06-31) happened more but April (Farvardin). It can be because of the Nowruz holiday in Iran. Fig. 21, indicates that *Akhbare Rasmi* has the least activity in *petroleum* news.

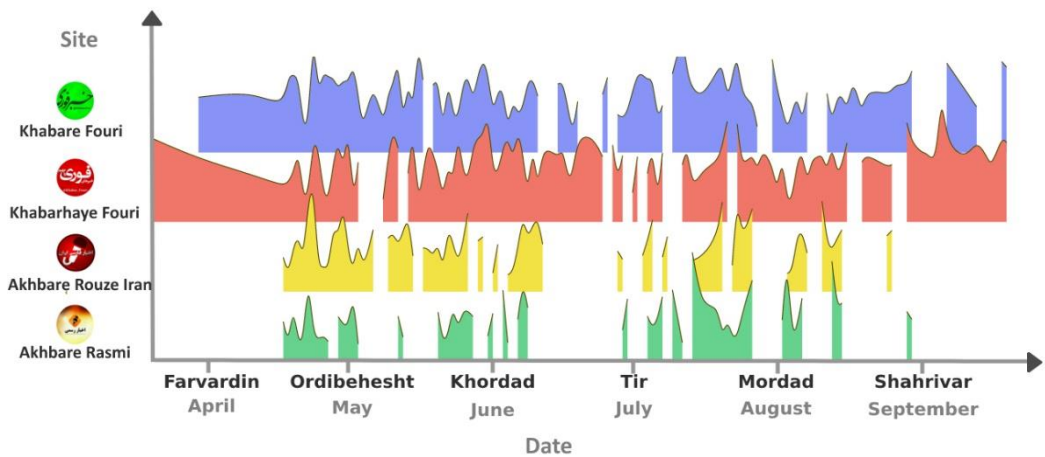

**Fig. 20. Justice shares news events in tempo-spatial dimensions in 2020 (1399).**

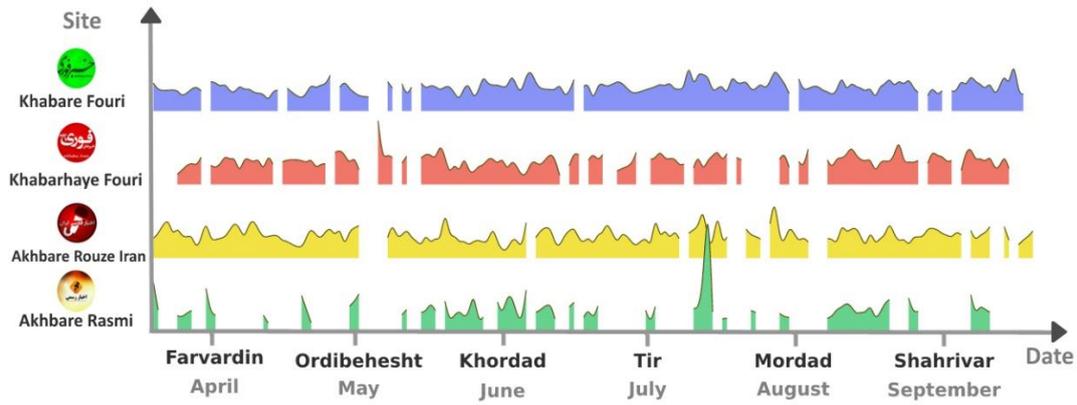
**Fig. 21. Petroleum news events in tempo-spatial dimensions in 2020 (1399).**

Fig. 22 demonstrates that in April (Farvardin) and July (Tir), the *quarantine* news is more and cause to stay at home on holiday.

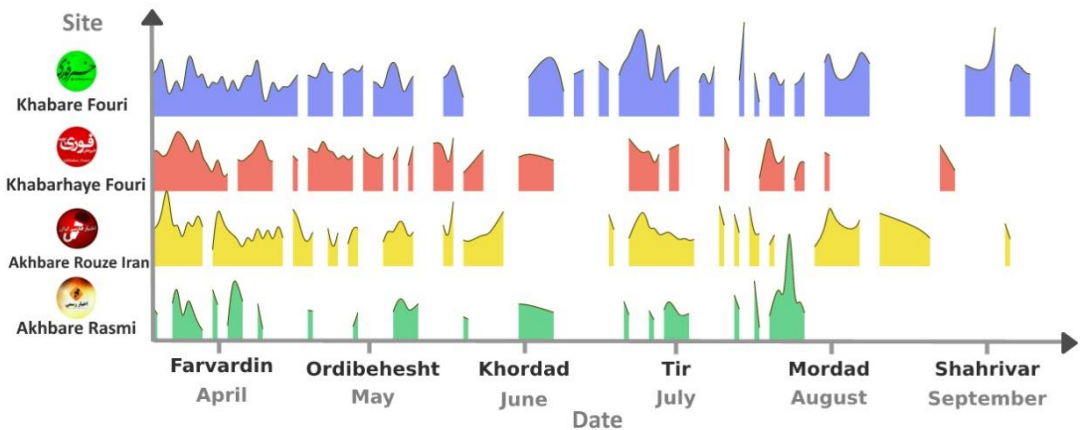
**Fig. 22. Quarantine news events in tempo-spatial dimensions in 2020 (1399).**

As summarized in Table 4, between March 23, 2020 (1399-01-01) and September 21, 2020 (1399-06-31), *Coronavirus (COVID-19)*, *vaccine*, and *flood* are the most frequent news events. The average of daily news about different events is presented in Table 5. It indicates that *Coronavirus (COVID-19)* news has attracted more attention in the pointed-out time interval.

**Table 4**
**The number of selected news events in the period from March 23, 2020 (1399-01-01), to September 21, 2020 (1399-06-31).**

|  | Khabare Fouri | Khabarhaye Fouri | Akhbare Rouze Iran | Akhbare Rasmi |
|---|---|---|---|---|
| **Coronavirus (COVID-19)** | 6850 | 6127 | 5357 | 3808 |
| **Vaccine** | 341 | 478 | 374 | 299 |
| **Reopening School** | 87 | 69 | 49 | 58 |
| **Earthquake** | 203 | 240 | 121 | 103 |
| **Fire** | 225 | 102 | 155 | 94 |
| **Flood** | 355 | 137 | 219 | 109 |
| **Justice shares** | 330 | 250 | 143 | 119 |
| **Petroleum** | 497 | 400 | 502 | 209 |
| **Quarantine** | 445 | 258 | 231 | 138 |

**Table 5**
**The daily average number of selected news events in the period from March 23, 2020 (1399-01-01), to September 21, 2020 (1399-06-31).**

|  | **Khabare Fouri** | **Khabarhaye Fouri** | **Akhbare Rouze Iran** | **Akhbare Rasmi** |
|---|---|---|---|---|
| **Coronavirus (COVID-19)** | 37 | 33 | 29 | 21 |
| **Vaccine** | 2 | 3 | 3 | 2 |
| **Reopening School** | 2 | 1 | 1 | 2 |
| **Earthquake** | 3 | 3 | 2 | 2 |
| **Fire** | 2 | 2 | 2 | 2 |
| **Flood** | 3 | 2 | 2 | 1 |
| **Justice shares** | 3 | 2 | 2 | 2 |
| **Petroleum** | 3 | 3 | 3 | 2 |
| **Quarantine** | 4 | 3 | 2 | 2 |

As explained in section 7, a system has been designed and developed that visualizes the correlation of Telegram archival data in the context of tempo-spatial as well as the meaning dimension. To achieve this, we need to extract the text, date, and time of each news entity from the Telegram news data archive. Then, by visualizing the extracted information, the correlation of the entities in the concept of time becomes visible. By searching via a keyword from the archival content, only the news related to that query phrase is seen. News entities are also depicted in a semantic dimension. The meaning dimension is the TF-IEF score for each news document. Some news entities that are chronologically correlated and have occurred almost the same time may also be semantically balanced.

## 8. Article achievements

This paper proposes a methodology for detailed investigation and modeling of data items regarding temporal, spatial, and tempo-spatial events and resources. Resolving the issues of data multiplications, diverse i18 coding and naming, and cross-platform implementation. Creating an efficient archival CDN that includes long-term and unique data with unified names is presented. By using the created archive in tempo-spatial CDN, temporal and spatial analytics and semantic correlation of events are possible.

On the other hand, one of the most important gains of the paper is the creation of a dataset from the Telegram news archive in JSON format that the scientific community and researchers can use in research contexts such as news archive analysis and NLP. This includes the features such as message text, time and date of the message, media files including video, photo, and audio, number of views, and so on.

## 9. Conclusions and Future Research

In this paper, an innovative strategy to effectively archive cyberspace data was presented. The archive is unified in tempo-spatial CDN. One of the prominent features of the TS-CDN is the management of files and duplicate media. In TS-CDN an efficient archive from Telegram at various time intervals is produced and then all data without interference and possible replacement of media files are unified. On the other hand, analysis and evaluation of Telegram data stream in the context of time and semantic is another aspect of this research. Data correlation in the time axis indicates the dependency and relevance of news entities at a particular point in time. In the spatial dimension, which includes diverse news sites and channels, the correlation of specific events based on the query is extracted from the Telegram data archive. The news content of the channels is monitored and tracked in a parallel approach from multiple cyberspace sources. Since the importance of news and its serious effects on society, the number of news sources can be increased for further research. Parallel processing and AI-annotated algorithms can also be used for boosting the speed of crawling processing and hash operation.

**Declaration of competing interest**

The authors declare that they have no known competing financial interests or personal relationships that could have appeared to influence the work reported in this paper.

**Acknowledgments**

This research did not receive any specific grant from funding agencies in the public, commercial, or not-for-profit sectors.